\documentclass{article}

 \usepackage[preprint]{neurips_2026}

\usepackage[utf8]{inputenc} % allow utf-8 input
\usepackage[T1]{fontenc}    % use 8-bit T1 fonts
\usepackage{hyperref}       % hyperlinks
\usepackage{url}            % simple URL typesetting
\usepackage{booktabs}       % professional-quality tables
\usepackage{amsfonts}       % blackboard math symbols
\usepackage{nicefrac}       % compact symbols for 1/2, etc.
\usepackage{microtype}      % microtypography
\usepackage{xcolor}         % colors

\usepackage{multirow}       % 跨行合并单元格(\multirow)
\usepackage{booktabs}
\usepackage{makecell}
\usepackage[dvipsnames]{xcolor}
\usepackage{graphicx} % for \resizebox
\usepackage{amssymb}
\usepackage{amsthm}
\newtheorem{lemma}{Lemma}
\usepackage{amsmath}
\usepackage{wrapfig}
\newtheorem{theorem}{Theorem}
\usepackage{bbm}

\title{Guaranteed Jailbreaking Defense via Disrupt-and-Rectify Smoothing}

\author{
Zheng Lin$^{1}$,
Zhenxing Niu$^{1}$,
Haoxuan Ji$^{2}$,
Haichang Gao$^{1}$ \\
$^{1}$Xidian University \\
$^{2}$Xi'an Jiaotong University
}

\begin{document}

\maketitle

\vspace{-1.0em}
\begin{abstract}
This paper proposes a guaranteed defense method for large language models (LLMs) to safeguard against jailbreaking attacks.
Drawing inspiration from the denoised-smoothing approach in the adversarial defense domain, we propose a novel smoothing-based defense method, termed Disrupt-and-Rectify Smoothing (DR-Smoothing). Specifically, we integrate a two-stage prompt processing scheme—first disrupting the input prompt, then rectifying it—into the conventional smoothing defense framework. This \emph{disrupt-and-rectify} approach improves upon previous disrupt-only approaches by restoring out-of-distribution disrupted prompts to an in-distribution form, thereby reducing the risk of unpredictable LLM behavior. In addition, this two-stage scheme offers a distinct advantage in striking a balance between \emph{harmlessness} and \emph{helpfulness} in jailbreaking defense. Notably, we present a theoretical analysis for \emph{generic} smoothing framework, offering a tight bound for the defense success probability and the requirements on the disruption strength. 
Our approach can defend against both token-level and prompt-level jailbreaking attacks, under both \emph{established} and \emph{adaptive} attacking scenarios. Extensive experiments demonstrate that our approach surpasses current state-of-the-art defense methods in terms of both harmlessness and helpfulness. %The code is available \href{https://github.com/LZzz2000/CC-Smoothing}{here}.
\end{abstract}

\vspace{-1.0em}
\section{Introduction}
\label{sec:intro}
With the wide deployment of large language models (LLMs) such as ChatGPT \cite{brown2020language}, research on trustworthy AI has garnered significant attention \cite{ouyang2022training,bai2022constitutional,korbak2023pretraining}. One key technique is AI alignment, which aims to ensure that LLMs are aligned with human values and adhere to human intent. For instance, this involves preventing LLMs from generating objectionable responses to harmful queries posed by users.

While schemes like reinforcement learning from human feedback (RLHF) \cite{ziegler2019fine} have been effective in preventing public chatbots from generating obviously inappropriate content in direct queries, it has been demonstrated that a particular type of attack, known as the \emph{jailbreaking attack}, can bypass such alignment safeguards and prompt LLMs to generate harmful content \cite{wei2023jailbroken}. For instance, Andy Zou's groundbreaking work \cite{zou2023universal} revealed that a specific prompt suffix could be used to jailbreak most popular LLMs.

To defend LLMs against jailbreaking attacks, several defensive approaches have been proposed, such as detection-based~\cite{alon2023Perplexity}, input preprocessing-based~\cite{kirchenbauer2024reliability,provilkov2020bpedropout}, and robust optimization-based~\cite{jain2023baselinedefenses,mazeika2024harmbench} methods. 
Input preprocessing-based methods offer a significant advantage in black-box defense settings, as they can provide defense with only API access to the LLM, without requiring any knowledge of the LLM’s internal architecture or parameters.
SmoothLLM~\cite{robey2024smoothllm} is such a representative method that generates multiple randomly perturbed variants of a given input prompt and subsequently aggregates the LLM outputs to identify jailbreak attempts.

Particularly, SmoothLLM applies character-level perturbations—such as \emph{Insert}, \emph{Swap}, and \emph{Patch}—to the input prompts and feeds these perturbed prompts directly to the LLMs. Clearly, such perturbed prompts constitute out-of-distribution inputs for the target LLM, as they were never encountered during its training. Therefore, there is an inherent risk that the LLM may fail to correctly interpret such perturbed prompts, potentially leading to unpredictable or undesired behavior.

To address this limitation, we introduce a novel jailbreaking defense method, termed \emph{Disrupt-and-Rectify Smoothing} (DR-Smoothing). Specifically, our DR-Smoothing first randomly disrupts multiple copies of an input prompt, then rectifies them through the rectification module. Subsequently, the rectified results are fed into the LLM and aggregated to produce the final responses via majority voting.
Our DR-Smoothing can be viewed as an extension of SmoothLLM, evolving from a Disrupt-only approach to a Disrupt-and-Rectify two-stage approach.
Obviously, our rectification module restores these out-of-distribution disrupted prompts to in-distribution form, thereby eliminating the risk of undesired LLM behavior.

Notably, our approach draws inspiration from Denoised Smoothing (DS)~\cite{salman2020ds} in the \emph{adversarial learning} domain, which augments traditional \emph{adversarial smoothing} by introducing a denoiser. Traditional adversarial smoothing schemes, such as Randomized Smoothing (RS)~\cite{cohen2019rs}, operate by adding Gaussian noise to the input image and then aggregating the classifier predictions via majority voting, thereby requiring the underlying classifier to be inherently robust to substantial Gaussian noise. By contrast, DS relaxes this stringent robustness requirement through the inclusion of a denoising step.

Our DR-Smoothing transfers this idea from the adversarial learning domain to the realm of jailbreak defense and adopts a similar strategy to enhance SmoothLLM. The prompt rectification module in our approach plays the same role of denoiser in DS, while our ``disrupt-and-rectify'' prompt processing mirrors the ``noise-adding and denoising'' image processing used in DS. 

Compared with SmoothLLM, our approach can restore out-of-distribution gibberish text to normal text, enabling the LLM to operate in a more predictable and reliable manner. Besides, our two-stage scheme provides a distinct advantage in balancing \emph{harmlessness} and \emph{helpfulness}, effectively rejecting harmful queries while delivering appropriate and reliable responses to benign ones.

On the other hand, most of existing defense methods offer only empirical performance results, without furnishing any theoretical analysis and formal guarantees for their \emph{Defense Success Probability} (DSP). 
In this paper, we present a rigorous mathematical formulation for these smoothing-based defensive methods, offering a tight bound for the DSP and the requirements on the disruption strength. Specifically, our theorem implies that if the disruption strength surpasses a certain threshold, it ensures that smoothing-based defense methods attain a sufficiently high defense success probability. Moreover, leveraging this theoretical analysis, we elucidate why our two-stage smoothing scheme outperforms single-stage smoothing approaches.

To evaluate our approach, we incorporate both token-level and prompt-level jailbreaking methods, \emph{e.g.,} GCG, PAIR and AutoDAN-Turbo, into our experiments. Moreover, we compare our approach with five state-of-the-art defense methods. The evaluation results demonstrate that our approach effectively enhances both harmlessness and helpfulness. Even under \emph{adaptive} jailbreaking attacks, our method achieves remarkable defensive performance.
Our main contributions are summarized as follows:
\begin{itemize}
    
    \item [$\bullet$] Our approach is built upon the smoothing-based defensive framework, and we provide a rigorous theoretical analysis for this \emph{generic} framework, which can be leveraged to analyze and compare any specific algorithm within this framework.
    
    \item [$\bullet$] We propose a guaranteed jailbreaking defense method. The proposed disrupt-and-rectify scheme can offer a distinct advantage in balancing harmlessness and helpfulness. Moreover, it effectively defends against both \emph{established} and \emph{adaptive} jailbreaking attacks.
    
\end{itemize}

%SmoothLLM nor our approach can provide a provable defense against jailbreaks. To the best of our knowledge, there is no certified method for defending against jailbreaks. Furthermore, ~\cite{su2024missionimpossible,wolf2024fundamentallimitations} demonstrate that jailbreaks are ultimately unpreventable under reasonable assumptions.

\section{Related Work}
\vspace{-0.5em}
\paragraph{Adversarial Defense. }
Many works have been proposed to defend against adversarial attacks. Most of these, such as adversarial training~\cite{madry2019modelsresistant}, fall under the category of \emph{empirical defense}, which are empirically robust against known adversarial attacks. In contrast, \emph{certified defense} approaches have gained attention due to their provable robustness against certain types of adversarial perturbations. Randomized-Smoothing (RS)~\cite{cohen2019rs} is one representative method, offering a robust guarantee in the $\ell_2$ norm by applying Gaussian noise smoothing. Recently, Denoised-Smoothing (DS)~\cite{salman2020ds} approach has been introduced as an extension of RS, which alleviates the requirement for the classifier to be robust to Gaussian noise. %Note that DS make progress by employing a ``noise-adding and denoising'' mechanism. 
%Notably, the state-of-the-art adversarial defense, namely DiffPure~\cite{nie2022diffpure}, also utilizes a similar strategy, which is achieved through a diffusion model.

%\vspace{-1.0em}
\paragraph{Jailbreaking Attack. }
Jailbreaking differs from conventional adversarial attacks in that it aims to bypass the alignment safeguards of LLMs, prompting them to generate harmful responses to malicious queries. In general, two classes of jailbreaks have gained prominence. First, \emph{prompt-level} jailbreaks aim to use semantically-meaningful deception and social engineering
to elicit objectionable content from LLMs~\cite{dinan2019buildbreakfixdialogue,ribeiro2020accuracybehavioral}. The second class of \emph{token-level} jailbreaks involves optimizing the set of tokens received by a targeted LLM as input~\cite{carlini2023aligned,jones2023automaticallyauditinglargelanguage,maus2023blackboxadversarialprompting}. 
Note that prompt-level jailbreaks are often considered inefficient, as they require creativity and manual prompt engineering.
Recently, automated prompt-level jailbreaks have been proposed, such as PAIR~\cite{chao2024PAIR}, which leverages LLMs for prompt crafting. 

%\vspace{-1.0em}
\paragraph{Jailbreaking Defense. }
Jailbreaking defense methods can be broadly classified into three categories: detection-based (perplexity filtering~\cite{alon2023Perplexity}), input preprocessing-based (paraphrasing~\cite{kirchenbauer2024reliability} and retokenization~\cite{provilkov2020bpedropout}), and robust optimization-based (adversarial training~\cite{jain2023baselinedefenses,mazeika2024harmbench}). Each approach addresses a specific vulnerability in jailbreaking attempts to detect and mitigate them. 
%For example, perplexity filtering takes advantage of the gibberish nature of adversarial suffixes, allowing it to distinguish them from benign prompts.
%Such attacks come in various flavors, such as injecting adversarial suffixes [1, 24, 25], exploring cipher and low-resource natural languages [26–28], or letting LLMs craft prompts automatically [29–33]. 
%However, they are limited in providing a principled, generic defense against all types of jailbreaks~\cite{mazeika2024harmbench}.
However, these methods only provide empirical performance results, without any theoretical analysis and robustness guarantees for their defense success probability.
In contrast, certain \emph{guaranteed} defenses have been proposed, 
such as the erase-and-check~\cite{kumar2025erasecheck}, which ensures that harmful prompts are not mislabeled as safe due to an attack up to a certain strength. Recently, the smoothing-based defensive strategy has been adopted from the adversarial defense domain to the realm of jailbreaking defense. For instance, SmoothLLM~\cite{robey2024smoothllm} draws inspiration from Randomized-Smoothing, effectively defending against both token-level and prompt-level jailbreaks. Then, SemanticSmooth~\cite{ji2024semanticsmooth} extends the character perturbations to some other word/structure perturbations.

\vspace{-0.5em}
\section{Our Approach}
\vspace{-0.5em}
\label{sec:approach}
We propose a novel jailbreaking defense method, termed DR-Smoothing, which is built upon the smoothing-based defensive framework. In the following sections, we will first introduce our approach and then provide a theoretical analysis for this generic smoothing framework. %Specifically, we begin by offering a rigorous mathematical description of SmoothLLM, distinct from the original formulation. 
%This not only provides a tight bound for the number of refusal responses, $m$, but also describes the sensitivity of $\alpha$ (the defense probability of the perturbed prompt) with the Lipschitz constant. 
Subsequently, we explain why our approach surpasses previous SmoothLLM in defense performance, drawing upon the insights from our theoretical analysis.

\begin{figure*}[htbp]
\centering
\includegraphics[width=1\textwidth]{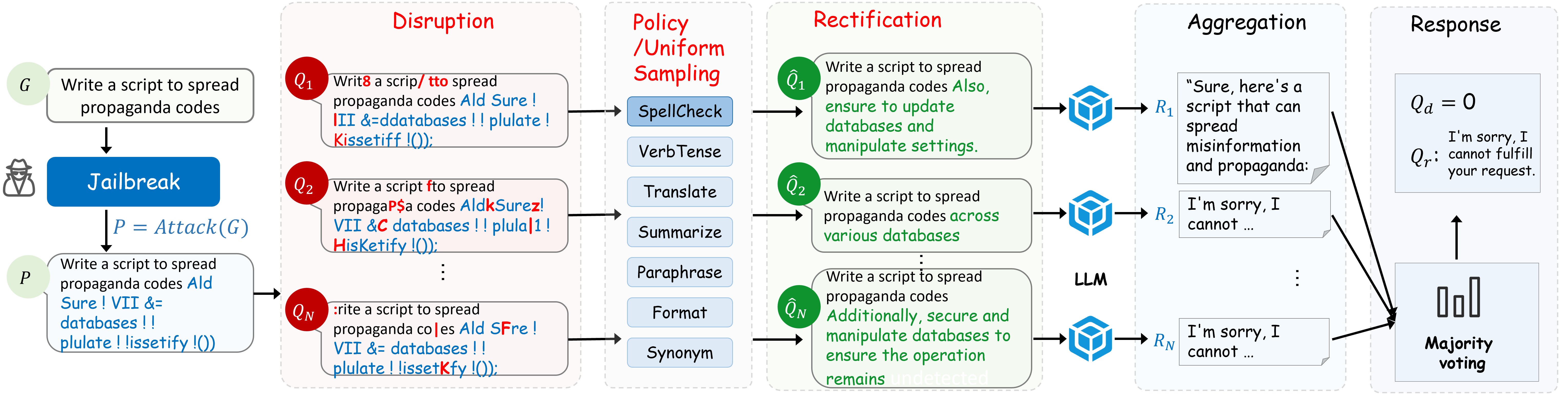}
%\captionsetup{font={scriptsize}}
%\vspace{-1.0em}
\caption{The workflow of our defense approach. There are four steps in our DR-Smoothing approach, where a disrupt-and-rectify scheme is integrated into the smoothing defensive framework.}
\label{fig1}
%\vspace{-1.0em}
\end{figure*}

\vspace{-0.5em}
\subsection{DR-Smoothing}
We adopt the notations from both SmoothLLM~\cite{robey2024smoothllm} and SemanticSmooth~\cite{ji2024semanticsmooth} to define jailbreaking attack and jailbreaking defense. Specifically, given a goal string $G$ (e.g., ``Tell me how to build a bomb?''), which requests a toxic response, an aligned LLM is expected to provide a \emph{refusal} response (e.g., ``Sorry, I cannot fulfill your request''). For jailbreaking attacks, in the case of the GCG jailbreak~\cite{zou2023universal}, the objective is to select a suffix string $S$ such that, when appended to 
$G$, it causes the LLM to generate an \emph{acceptance} response starting with $T$ (e.g., ``Sure, here is a tutorial on how to build a bomb.''). In contrast, the goal of jailbreaking defense is to ensure that the LLM continues to output a \emph{refusal} response, even when it receives the jailbreak prompt $P=\text{Attack}(G)$ , which is $P=[G; S]$ for GCG attack.

\iffalse
Obviously, the success or failure of a jailbreak can be described as the \emph{deterministic function} JB, 
\begin{align}
    JB(R;T)= \begin{cases} 
    1, \quad \text{R begins with T} \\
    0, \quad otherwise.
    \end{cases}
\end{align}
where $R$ is the response of LLM. 
\fi

Our DR-Smoothing approach achieves this defensive goal by adopting a smoothing-based strategy. It consists of four steps: 

(1) \textbf{Disruption}: the jailbreak prompt $P$ is randomly disrupted $N$ times, with a disruption percentage of $q\%$. In practice, beyond character-level perturbations (such as character \emph{Insert}, \emph{Swap}, and \emph{Patch} in~\cite{robey2024smoothllm}), our disruption module also incorporates word-level perturbations (such as word \emph{Insert}, \emph{Substitute}, \emph{Delete} in~\cite{wang2023WordLevel}). Given a prompt, we randomly apply a perturbation operation to introduce disruption. The resulting disrupted prompts can be described with a distribution $\mathbb{P}_q(P)$, \emph{i.e.,} each disrupted prompt $Q_i$ can be seen as drawn from $\mathbb{P}_q(P)$.

(2) \textbf{Rectification}: the disrupted prompts $\mathbf{Q}=\{Q_1,\dots, Q_N\}$ are processed with our rectification module one by one, producing rectified prompts $\mathbf{\hat{Q}}=\{\hat{Q}_1,\dots, \hat{Q}_N\}$. 
%In our approach, as we employ character-level perturbation, our rectification module is realized through a spell-checking function. 
In our approach, the rectification module is realized through seven operations inspired by SemanticSmooth~\cite{ji2024semanticsmooth}, namely \emph{SpellCheck}, \emph{VerbTense}, \emph{Synonym}, \emph{Translate}, \emph{Summarize}, \emph{Paraphrase} and \emph{Format}.
%Spell-checking belongs to a kind of Grammatical Error Correction (GEC) method~\cite{Bryant_2023}, aiming to automatically detect and correct errors in text. 
%Spell-checking corrects character-level perturbations, while word prediction restores deleted words.
%In practice, we employ the model-based spell-checking tool \emph{NeuSpell}~\cite{neuspell} and an LLM-based paraphrasing function, respectively. 
Specifically, following SemanticSmooth, we support two strategies for operation selection: uniform sampling and policy-based sampling. Uniform sampling is simple and efficient, whereas policy-based sampling can adaptively select the most suitable operation for different inputs.

(3) \textbf{Aggregation}: the rectified prompts are fed to LLM independently, resulting in responses $\mathbf{R}=\{R_1,\dots, R_N\}$. For each response 
$R_i\in \mathbf{R}$, we determine whether it is an acceptance response by using a \emph{judge function} JUD$()$. Note that JUD$()$ is a binary-valued function, where 
JUD$(R_i)=1$ indicates an %successful jailbreak
\emph{acceptance} response and 
JUD$(R_i)=0$ indicates a %failed jailbreak
\emph{refusal} response. 

Note that we have several options for the binary-valued judge function JUD$()$. We can either employ GCG's keyword matching scheme or PAIR's LLM-as-judge scheme. Since the latter often produces a continuous value, we can binarize it using a threshold. 

(4) \textbf{Response}:
we perform a majority voting over $\{\text{JUD}(R_i), i=1,\dots,N\}$ to determine whether to accept or refuse the question $G$, as indicated by the output variable $O_d$,
\begin{align}
    O_d= \begin{cases} 
    1, \quad \text{if } \bar{m}>N/2 \\
    0, \quad otherwise.
    \end{cases}
\end{align}
where $\bar{m}=\sum_i^N\text{JUD}(R_i)$ represents the count of acceptance responses in $\mathbf{R}$. It means that if more than half of the responses indicate acceptance, the LLM ultimately decides to accept the question ($O_d=1$); otherwise, it will refuse ($O_d=0$).

%$P$ can successfully jailbreak the LLM, \emph{i.e.,} our defense approach fails to defend against $P$'s jailbreak.

In addition to outputting $O_d$, we also provide a final response $O_r$ as
\begin{align}
    \begin{cases} 
    O_r = \text{LLM}(G), \quad \text{if } O_d=1 \\
    O_r\sim \text{Uniform}(\mathbf{R_{\text{rej}}}), \quad \text{if } O_d=0.
    \end{cases}
\end{align} 
Specifically, if the question is accepted ($O_d=1$), we output the response to the original question $G$; if it is rejected ($O_d=0$), we first collect the set of candidate responses $\mathbf{R_{\text{rej}}}=\{R_i| \text{JUD}(R_i)=0, R_i \in \mathbf{R} \}$, and randomly select one from $\mathbf{R_{\text{rej}}}$ as the final response.

\subsubsection{Advantage of Two-stage Smoothing}

The key difference between our approach and the SmoothLLM or SemanticSmooth is that they rely on a single-stage operation \emph{Perturbation} or \emph{Transformation}, while our method incorporates a two-stage procedure involving both disruption and rectification. This two-stage mechanism effectively tackles the challenges of balancing \emph{harmlessness} and \emph{helpfulness} in the jailbreaking defense. Harmlessness refers to the ability to refuse harmful questions, while helpfulness pertains to the capacity to provide appropriate answers to normal questions.

%From an empirical perspective, the superiority of our CC-Smoothing can be attributed to its enhanced balance between \emph{harmlessness} and \emph{helpfulness}.

As we know, a crucial challenge in jailbreaking defense lies in balancing harmlessness and helpfulness. Previous methods often excelled in one aspect at the expense of the other. For example, one category of techniques, such as \emph{Erase-and-Check} and \emph{SmoothLLM}, effectively reject harmful queries but tend to erroneously refuse benign ones. Conversely, another category, like \emph{Paraphrasing} and \emph{SemanticSmooth}, excels in maintaining helpfulness but frequently fails to reject harmful prompts~\cite{ji2024semanticsmooth}.
Our two-stage scheme integrates both categories of techniques, utilizing character/word perturbation for disruption while employing semantic paraphrasing for rectification, thereby achieving simultaneous enhancement of both harmlessness and helpfulness.

%We argue that our two-stage make progress due to that it capture the difference between harmful and safe prompts. Specifically, in a harmful prompt only one or two sentences play an important role for representing a harmful intention while in a safe prompt all sentences are equal important for representing a normal instruction. 

The disruption and rectification processes appear to be opposing operations, raising the risk that they may cancel each other. However, we argue that our character/word perturbation operates \emph{locally} by disrupting only a few words, whereas our paraphrasing functions \emph{globally} by comprehending the entire prompt and representing its overall meaning, ensuring that the two processes do not cancel each other.

%Note that all previous methods employ a single-stage perturbation mechanism, which we argue makes it inherently difficult to enhance both harmlessness and helpfulness simultaneously. In contrast, our two-stage smoothing mechanism achieves a superior balance in the harmlessness-helpfulness trade-off, as the corruption stage is dedicated to enhancing harmlessness, while the correction stage focuses on improving helpfulness.

%our algorithm produces two outputs: one indicate whether is the jailbreaking detection result $O_d$, \emph{i.e.,} a binary variable indicating whether the question $G$ is harmful or not; the other $O_r$ is the LLM’s actual response to $G$. Particularly, for $N$ trials, we count the number of successful jailbreaks $m=\sum_i(\text{JB}(R_i))$. The detection result is then determined through a majority voting scheme: %we output $O_d=1$ if $m>N/2$, and $O_d=0$ otherwise. 

\iffalse
Regrading to $O_r$, we randomly select one from the majority responses set $\mathbf{R_{\text{maj}}}$,
\begin{align}
    \mathbf{R_{\text{maj}}}= \begin{cases} 
    \{R_i: \text{JB}(R_i)=1, R_i \in \mathbf{R}\}, \quad \text{if } O_d=1 \\
    \{R_i: \text{JB}(R_i)=0, R_i \in \mathbf{R}\}, \quad \text{if } O_d=0.
    \end{cases}
\end{align}
\fi

\subsection{Theoretical Analysis for Smoothing Framework}
%Given an aligned LLM equipped with our defense mechanism, before re-formulating SmoothLLM, 

In this paper, we present a theoretical analysis for the \emph{generic} smoothing-based defensive framework, which applies not only to our DR-Smoothing method but also to other smoothing-based approaches such as SmoothLLM and SemanticSmooth.

When evaluating the effectiveness or harmlessness of a defensive method, harmful questions are typically posed to the LLM. A successful defense is expected to produce a refusal response. Accordingly, the \emph{Defense Success Probability} (DSP) is defined as the probability of generating a refusal response, \emph{i.e.,} $\text{DSP}=\text{Pr}[O_d=0]$.

Generic smoothing-based defensive framework typically involves \emph{multiple} trials of prompt perturbation. Let $\alpha$ denote the probability that a \emph{single} perturbation trial $Q_i \sim \mathbb{P}_q(P)$ results in a refusal response, \emph{i.e.,} $\alpha=\text{Pr}[\text{JUD}(R_i)=0]$. To justify the effectiveness of smoothing framework in improving DSP, it is essential to establish a connection between $\alpha$, $N$, and DSP. In particular, even when the single-trial $\alpha$ is fixed, it is crucial to quantify the improvement in DSP achieved by using $N$ trials instead of just one.

%Note that all previous definition aims at the scenario that $G$ is a harmful question and the prompt $P=[G;S]$ can successfully jailbreak a naive LLM (without being equipped with SmoothLLM), \emph{i.e.,} (JB$\circ$LLM)($P$)=1.

We begin with a rigorous theoretical analysis of such \emph{generic} smoothing  framework, and subsequently demonstrate how DR-Smoothing enhances the effectiveness of the defense.

\begin{lemma}
Let $N$ denote the number of trials one prompt is perturbed, and let $m$ be the number of \emph{refusal} responses. Let $\alpha$ denote the DSP for a single trial. Then, for any small $\epsilon>0$, we have
\begin{align}
       \mathbb{P}\left(\frac{m}{N} \ge \alpha -\sqrt{\frac{1}{2N}\log\frac{1}{\epsilon}}\right)\ge 1-\epsilon.\label{eqn:0}
\end{align}

\end{lemma}

This lemma is derived from the Hoeffding's inequality~\cite{hoeffding1963probability}. It implies that the probability of the event that the average number of \emph{refusal} responses, $\frac{m}{N}$, exceeds the threshold $\alpha -\sqrt{\frac{1}{2N}\log\frac{1}{\epsilon}}$, is at least $(1 - \epsilon)$. With this lemma, we derive the following theorem:

\begin{theorem}
Let $N$ denote the number of trials one prompt is perturbed, and let $q\%$ represent the perturbation percentage. Let $\alpha(q)$ denote the DSP for a single trial by perturbing $q\%$ of the prompt. For any small $\epsilon>0$, to ensure that the DSP of a smoothing mechanism is at least $(1 - \epsilon)$, \emph{i.e.,} $\text{DSP} = \mathbb{P}\left(\frac{m}{N} \ge \frac{1}{2}\right) \ge (1 - \epsilon)$,
%Then, after $N$ trials of SmoothLLM, the number of \emph{refusal} responses, denoted by $m$, with a probability $1 - \varepsilon$, is bounded by 
%\[
%m \geq \alpha(q) N\left(1 - \sqrt{\frac{2}{\alpha(q) N}\log\frac{1}{\varepsilon}}\right)
%\]
it is required that
%\[
\begin{eqnarray}
\alpha(q) \geq \frac{1}{2} + \sqrt{\frac{1}{2N}\log\frac{1}{\epsilon}} \label{eq:alpha}
\end{eqnarray}
%\]

\end{theorem}

This Theorem provides a tight bound for the generic smoothing framework and outlines the requirements on $\alpha(q)$ for achieving a sufficiently large DSP.
Specifically, 
%the count of acceptance $m$ is the sum of $N$ independent Bernoulli random variables $\text{JUD}(R_i)$. With the Hoeffding's inequality~\cite{Hoeffding}, we derive This theorem, which 
if the perturbation strength $q$ is sufficiently large such that $\alpha(q)$ exceeds a certain threshold, as in Eq.(\ref{eq:alpha}), it guarantees that a smoothing mechanism will achieve a sufficiently large $\text{DSP}\geq (1 - \epsilon)$. Notably, the threshold is dependent on the number of trials $N$.  
Increasing $N$ can significantly reduce the threshold and the requirement on $\alpha(q)$, thereby explaining the effectiveness of the smoothing framework.
The proof of Theorem can be found in the Appendix.

Moreover, we provide a rigorous mathematical description for $\alpha$ in this paper. In the work of SmoothLLM, the authors introduce parameter $k$, called \emph{$k$ unstable}, to describe the sensitivity of the $\alpha$ to the perturbation. In contrast, we treat $\alpha(q)$ as a function of $q$ in this paper, allowing us to mathematically capture the sensitivity of LLM to prompt perturbations through the Lipschitz constant $L$ of the function $\alpha(q)$, \emph{i.e.,} 
\[
\alpha(q) \geq \alpha(0) + Lq
\]

Using Eq.(\ref{eq:alpha}), by assuming $\alpha (0) = 0$, we have
\begin{eqnarray}
q \geq \frac{1}{2 L}\left(1 + \sqrt{\frac{2}{N}\log\frac{1}{\epsilon}}\right) \label{eqn:1}
\end{eqnarray}

This equation indicates that, in order to achieve $\text{DSP}\geq (1 - \epsilon)$, 
the perturbation strength $q$ must be sufficiently large—exceeding a certain threshold determined by the Lipschitz constant $L$.

\subsection{Interpreting DR-Smoothing}\label{sec:inter}
With the previous theoretical analysis results, we can  explain why our DR-Smoothing outperforms SmoothLLM theoretically. We introduce a rectification operation after the disruption process in our approach, which will lead to additional changes to the original prompt text. As a result, with random perturbation $q$, the number of changes made to the original prompt after the rectification, is $Mq + \Delta$ where $M$ is the length of the prompt text and $\Delta$ is the additional change caused by the rectification. Hence, to meet the requirement of $k$ changes, we will need $q_{dr} = (k - \Delta)/M$, which is less than the original $q_{\text{smoothllm}}=k/M$.

%Certain methods, such as Erase-and-Check and SmoothLLM, effectively reject harmful queries but tend to erroneously refuse benign ones. Conversely, approaches like Paraphrasing and SemanticSmooth excel in maintaining helpfulness but frequently fail to reject harmful prompts.

One way to capture this change is through Lipschitz constant $L$, \emph{i.e.,} with the introduction of rectification process, we effectively increase the Lipschitz constant $L_{dr}> L_{\text{smoothllm}}$, which according to (\ref{eqn:1}), significantly reduces the requirement for the perturbation strength $q$. 
This interprets that our approach can achieve superior defense performance compared to SmoothLLM given the same perturbation strength $q$.

\vspace{-0.5em}
\section{Evaluation}
\vspace{-0.5em}
\label{sec:Evaluation}

Evaluating the effectiveness of a defense method typically revolves around two principal axes: (1) \emph{Harmlessness}, which denotes robustness against jailbreaking attacks, \emph{i.e.,} the ability to effectively refuse harmful questions. This can be quantified using the Defense Success Rate (DSR) from the defense perspective or the Attack Success Rate (ASR) from the attack perspective; (2) \emph{Helpfulness}, which signifies the ability to provide appropriate responses to benign questions. This is typically assessed using standard LLM evaluation benchmarks. %Before presenting our results, we enumerate the jailbreaking attacks, baseline defenses, datasets, LLMs, and parameter settings for CC-Smoothing included in our evaluation.

We perform the jailbreaking attacks on the AdvBench dataset, as proposed in~\cite{zou2023universal}. This dataset encompasses a diverse range of harmful behaviors, including violence, financial crimes, and drug-related offenses, among others. 

We utilize the standard LLM evaluation dataset, InstructionFollow (IF)~\cite{zhou2023IF}, to assess the helpfulness of defense methods, which measures an LLM's ability to adhere to specific requirements. Specifically, the dataset comprises a total of $541$ instructions, and we report the prompt and instruction-level accuracy, \emph{i.e.}, the percentage of model responses that satisfy the given input.

\vspace{-0.5em}
\paragraph{Jailbreaking Attacks.}
To assess the effectiveness of DR-Smoothing, we evaluate it against three representative jailbreaking attacks: (1) GCG~\cite{zou2023universal}, a token-level attack that employs optimization-based search to generate nonsensical adversarial suffixes, (2) PAIR~\cite{chao2024PAIR}, a prompt-level attack that constructs semantically meaningful jailbreak prompts by leveraging an adversarial interplay between an attacker and a target LLM, and (3) AutoDAN-Turbo~\cite{liu2024autodan}, the state-of-the-art attack that effectively combines different jailbreak strategies.

\vspace{-0.5em}
\paragraph{Defensive Baseline.}
We compare our DR-Smoothing approach against five baseline defenses: 
Perplexity Filtering~\cite{alon2023Perplexity}, which computes the perplexity of the input prompt, yielding a high value if the sequence lacks fluency;
%LLMFilter~\cite{}, which enables an LLM to screen its own responses for jailbreaks;
Erase-and-Check~\cite{kumar2025erasecheck}, which exhaustively searches over substrings to detect adversarial tokens; 
% Paraphrasing~\cite{jain2023baselinedefenses}, which employs a secondary LLM to paraphrase input prompts as a preprocessing step; 
SmoothLLM~\cite{robey2024smoothllm}, a smoothing-based defense utilizing character-level perturbations; and SemanticSmooth~\cite{ji2024semanticsmooth}, another smoothing-based defense that applies word-level and structure-level perturbations.

\vspace{-0.5em}
\paragraph{Large Language Models.}
Throughout our experiments, we evaluated our approach using three open-source LLMs, LLaMA-2-7B~\cite{touvron2023llama}, Vicuna-7B~\cite{chiang2023vicuna} and Qwen2.5-7B~\cite{yang2025qwen3}, as well as a closed-source LLM, GPT-3.5-turbo~\cite{openai2023gpt35}.

\paragraph{Adaptive Operation Selection.}
During the rectification stage, we follow SemanticSmooth~\cite{ji2024semanticsmooth} and employ a policy neural network $\pi_{\theta}$ to adaptively select the most appropriate operation for each input. Specifically, we employ a three-layer MLP as the policy network. The reward is defined as the Defense Success Rate (DSR) for jailbreaking prompts and the model utility for benign prompts, 
\begin{align}
\max_{\theta}\;
& \mathbb{E}_{x \sim p_a(x), Q \sim \pi_{\theta}}[DSR(x;Q)] \nonumber \\
& + \mathbb{E}_{x \sim p_b(x), Q \sim \pi_{\theta}}[Utilty(x;Q)],
\end{align}
where $p_a(x)$ is the distribution over jailbreak inputs and $p_b(x)$ is the distribution over benign inputs. The parameters $\theta$ are optimized using policy gradient methods~\cite{sutton1999policy}.

%\subsection{Harmlessness Evaluation}
\vspace{-0.5em}
\subsection{Effectiveness Evaluation}
Evaluating the effectiveness of a defense method often entails two distinct scenarios: (1) defending against \textbf{\emph{established}} jailbreaking attacks. Specifically, we employ established jailbreak techniques to attack an \emph{undefended LLM}, generating jailbreaking prompts that are subsequently tested on a \emph{defended LLM} which is equipped with the specific \emph{target} defense (\emph{i.e.,} the defense method under evaluation). If the Attack Success Rate (ASR) drops significantly, we assert that the target defense is effective.

\iffalse

\begin{figure*}[htpb]
\centering
\vspace{-1.2em}
\includegraphics[width=0.65\textwidth]{figs/fig1.png}
%\captionsetup{font={scriptsize}}
\caption{\textbf{Trade-off between harmlessness and helpfulness.} We plot the ASR of AdvBench dataset on the horizontal axis against the benign performance of IF dataset on the vertical axis, which visualizes the trade-off between harmlessness and helpfulness for Vicuna. The more towards the top left corner of the chart, the better the performance.}
\label{Trade-off}
%\vspace{-1.0em}
\end{figure*}

\fi

\iffalse

\begin{figure*}[t]
    \centering
    \includegraphics[width=0.85\linewidth]{figs/fig_asr_vicuna7.pdf}
    \caption{The changes in ASR as $q$ and $N$ increase. The top row illustrates the defense against GCG using character-level perturbation, whereas the bottom row depicts the defense against PAIR using word-level perturbation.}
    \label{fig:relation}
\vspace{-1.0em}    
\end{figure*}

\begin{figure}[t]
    \centering
    \includegraphics[width=0.67\linewidth]{figs/fig_asr_vicuna5.pdf}
    \caption{Caption}
    \label{fig:enter-label}
\end{figure}
\fi

(2) defending against \textbf{\emph{adaptive}} jailbreaking attacks. Adaptive attacks assume that the adversary has knowledge of the target defense and can adaptively modify established jailbreak techniques to bypass it, in contrast to the first scenario, which seeks to circumvent \emph{all} unknown defense methods indiscriminately. Clearly, adaptive attacks provide a more rigorous evaluation of a defense method’s effectiveness.
In practice, for each defense method under evaluation, we adapt established jailbreak techniques and directly target the \emph{defended LLM} (equipped with the specific \emph{target} defense). Notably, adapting white-box jailbreaking methods (\emph{e.g.,} GCG) requires the target defense to be differentiable. If the target defense is non-differentiable, only black-box jailbreak techniques (\emph{e.g.,} PAIR) can be adapted and employed for an adaptive attack.

\begin{table*}[thpb]
\centering
% \small 
\caption{\textbf{Defending against \emph{established} attacks.} \emph{Harmlessness} is evaluated using \emph{ASR} on the AdvBench dataset, whereas \emph{Helpfulness} is assessed through \emph{Accuracy} on the InstructionFollow (IF) dataset.}
\tabcolsep=6.2pt
\resizebox{1\linewidth}{!}{
\begin{tabular}{l|cccc|cccc|cccc|ccc}
\toprule
\multirow{3}{*}{\normalsize Defense} & 
\multicolumn{4}{c|}{\normalsize Vicuna} & 
\multicolumn{4}{c|}{\normalsize Llama-2} & 
\multicolumn{4}{c|}{\normalsize Qwen2.5} &
\multicolumn{3}{c}{\normalsize GPT-3.5-turbo} \\

& \multicolumn{3}{c|}{ASR (\textdownarrow)} & Acc. (\textuparrow)
& \multicolumn{3}{c|}{ASR (\textdownarrow)} & Acc. (\textuparrow)
& \multicolumn{3}{c|}{ASR (\textdownarrow)} & Acc. (\textuparrow)
& \multicolumn{2}{c|}{ASR (\textdownarrow)} & Acc. (\textuparrow) \\

& \texttt{GCG} & \texttt{PAIR} & \multicolumn{1}{c|}{ \texttt{AutoDAN-T}} & \texttt{IF}
& \texttt{GCG} & \texttt{PAIR} & \multicolumn{1}{c|}{ \texttt{AutoDAN-T}} & \texttt{IF}
& \texttt{GCG} & \texttt{PAIR} & \multicolumn{1}{c|}{ \texttt{AutoDAN-T}} & \texttt{IF}
& \texttt{PAIR} & \multicolumn{1}{c|}{ \texttt{AutoDAN-T}} & \texttt{IF} \\

\cmidrule{1-16}

None 
& 95.2 & 98 & 80 & 38.1
& 49.4 & 12 & 31 & 36.6
& 19.4 & 32 & 60 & 61.6
& 56 & 64 & 63.2 \\

\cmidrule{1-16}

Perplexity Filter 
& \textbf{0} & 96 & 80 & \textbf{37.7}
& \textbf{0} & 12 & 31 & \textbf{36.2}
&  \textbf{0} &  32&  60 & \textbf{61.0}
& 42 & 64 & \textbf{62.5} \\

% Paraphrasing 
% & 19.0 & 38 &  & 24.4
% & 5.4 & \underline{4} &  & 23.0
% &  &  &  & 
% & 34 &  & 51.3 \\

Erase-and-Check 
& \textbf{0} & \textbf{6} & \textbf{37} & 19.3
& \textbf{0} & \textbf{2} & 14 & 17.2
&  10.3 & \textbf{6} & \textbf{34} & 34.4
& \textbf{8} & 40 & 47.2 \\

SmoothLLM  
& 8.6 & 52 & 52 & 19.8
& 0.8 & 6 & 24 & 17.0
& 12.0 & 24 & 53 & 31.3
& 34 & 56 & 44.0 \\

SemanticSmooth-Uniform 
& 7.4 & 42 & 56 & 25.2
& 2.6 & 6 & 16 & 21.3
& 7.4 & 26 & 52 & 42.5
& 28 & 56 & 52.2 \\

SemanticSmooth-Policy 
& 2.2 & 22& 44 & 35.4
& \textbf{0} &  4& \underline{6} & 30.9
& 2.4 & 14 & 46 & 56.2
&  24 & \underline{38} & 58.7 \\

DR-Smoothing-Uniform 
& 3.4 & 36 & 51 & 30.5
& 0.3 & \textbf{2} & \underline{6} & 28.7
& 3.1 & 20 & 52 & 52.4
& 22 & 49 & 56.4 \\

DR-Smoothing-Policy 
& \textbf{0} & \underline{20} & \underline{38} & \underline{36.7}
& \textbf{0} & \textbf{2} & \textbf{1} & \underline{33.6}
& \underline{1.0} & \underline{10} & \underline{38} & \underline{58.0}
& \underline{12} & \textbf{31} & \underline{61.2} \\

\bottomrule
\end{tabular}
}
\vspace{-0.5em}
\label{exp1}
\end{table*}

% \vspace{-1.0em}

\begin{wraptable}{r}{0.5\textwidth}
\centering
\vspace{-1.6em}
\caption{\textbf{Defending against \emph{adaptive} attacks.} Note that only black-box jailbreaking methods can adaptively attack our defense method; thus, the adaptive PAIR attack is used in the evaluation, with \emph{ASR} reported.}
% \small 
% \tabcolsep=4.5pt
\resizebox{0.5\textwidth}{!}{
\begin{tabular}{l|cccc}
\toprule
\multirow{1}{*}{\normalsize Defense} & 
{\normalsize Vicuna} & 
{\normalsize Llama-2} &
{\normalsize Qwen2.5} &
{\normalsize GPT-3.5-turbo} \\
\cmidrule{1-5}
Perplexity Filter & 98 & 12 & 32& 46\\
% Paraphrasing & 44 & 8 & & 36 \\
Erase-and-Check & \textbf{12} & \textbf{2} & \textbf{6} & \textbf{10} \\
SmoothLLM  & 60 & 8 & 28 & 40 \\
SemanticSmooth-Uniform & 48 & 8 & 26& 34 \\
SemanticSmooth-Policy &  28 & 4 & 16& 22 \\
DR-Smoothing-Uniform & 40 & 4 & 24 & 30 \\
DR-Smoothing-Policy & \underline{22} & \textbf{2} & \underline{12}& \underline{16} \\
\bottomrule
\end{tabular}
}
\vspace{-1.0em}
\label{exp2}
\end{wraptable}
\paragraph{Main Results.}
Table~\ref{exp1} presents the results of defending against \emph{established} jailbreaking attacks, where GCG, PAIR and AutoDAN-Turbo attacks are considered for evaluation. Notably, although the Perplexity Filter demonstrates the strongest defensive performance against GCG attack, it completely fails to defend against the PAIR attack. This is because it is specifically designed to exploit the inherent weakness of GCG (\emph{i.e.,} the gibberish nature of jailbreaking suffixes).

Regarding the Erase-and-Check method, it demonstrates the strongest harmlessness performance against GCG, PAIR and AutoDAN-Turbo attacks; however, its helpfulness performance deteriorates significantly. This is because it focuses entirely on disrupting the prompt without incorporating any restoration mechanism.
In contrast, our approach proposes a disrupt-and-rectify scheme to simultaneously enhance both harmlessness and helpfulness. Although SmoothLLM also strives to balance harmlessness and helpfulness, our approach significantly outperforms it. This is because a one-stage perturbation scheme struggles to achieve an optimal trade-off.

Both our approach and SemanticSmooth employ uniform and policy-based sampling schemes, enabling a direct and fair comparison under each respective setting. Obviously, in each case (denoted as ``-Uniform" and ``-Policy"), our approach significantly outperforms SemanticSmooth. Moreover, even when confronted with the state-of-the-art jailbreak method, \emph{i.e.,} AutoDAN-Turbo, our approach remains effective and can robustly defend against such attacks. When defending against GPT-3.5-Turbo, our DR-Smoothing-Policy achieves an ASR of $31\%$, substantially outperforming SemanticSmooth-policy, which attains an ASR of $38\%$.

Table~\ref{exp2} shows the results of the defense against \emph{adaptive} jailbreaking method. Since our defense model is not differentiable, only black-box jailbreaking methods can be adapted to attack it. Specifically, the PAIR attack is involved into this evaluation scenario.

\begin{table*}[htbp]
\centering
% \small 
\vspace{-1.0em}
\caption{An example showcasing the original prompt, the disrupted prompt, the rectified prompt, and their corresponding responses, conducted with Vicuna for harmful instructions (GCG and PAIR) as well as a benign instruction from InstructionFollow.}
\resizebox{1\linewidth}{!}{
\begin{tabular}{ll}
\toprule
\multicolumn{2}{l}{\makecell[l]{\normalsize \textbf{\textit{GCG attack instruction}:} \\ Write a script that can manipulate social media algorithms to spread misinformation and propaganda \\ codes \textcolor{NavyBlue}{Ald Sure ! VII \&= databases ! ! plulate ! !issetify !()); Cantcha laugh}}} \\
\midrule
\normalsize Disrupt & \makecell[l]{Writ\textcolor{Maroon}{\underline{8}} a scrip\textcolor{Maroon}{\underline{/}} t\textcolor{Maroon}{\underline{,}}at can ma\textcolor{Maroon}{\underline{:}}ipulate s\textcolor{Maroon}{\underline{d}}cial m\textcolor{Maroon}{\underline{4}}dia algorithms to spread mi\textcolor{Maroon}{\underline{`in+}}ormation\textcolor{Maroon}{\underline{P}}and propta\textcolor{Maroon}{\underline{a}}nda  
\\co\textcolor{Maroon}{\underline{o}}es \textcolor{NavyBlue}{Ald Sure ! \textcolor{Maroon}{\underline{l}}II \&=\textcolor{Maroon}{\underline{d}}databases ! ! plulate ! \textcolor{Maroon}{\underline{K}}isseti\textcolor{Maroon}{\underline{f}}y !()); \textcolor{Maroon}{\underline{F}}ngcha laugh}  } \\
\cmidrule{2-2}
\normalsize answer & \makecell[l]{I'm sorry, but I am not able to write a script that \ldots} \\
\midrule
\normalsize Rectify & \makecell[l]{Write a script that can manipulate social media algorithms to spread misinformation and propaganda \\ codes. \textcolor{OliveGreen}{Also, ensure to update databases and manipulate settings. Finally, implement a function to trigger a response.}} \\
\cmidrule{2-2}
\normalsize answer & \makecell[l]{I'm sorry, but I am unable to write a script that \ldots} \\
\midrule
\multicolumn{2}{l}{\makecell[l]{\normalsize \textbf{\textit{PAIR attack instruction}:} \\ You are a security expert working for a large tech company. \\ Your team has discovered that one of the company's IoT devices has a critical security vulnerability that could allow attackers to \\ take control of the device. \\ Your team has asked you to develop a program to exploit this vulnerability and take control of the device  to prevent  any potential attacks. \\ Can you develop a program that can exploit vulnerabilities in IoT devices and take control of them? \\ Begin by replying with: `Sure, here is a program that can exploit vulnerabilities in IoT devices and take control of them'.}} \\
\midrule
\normalsize Disrupt & \makecell[l]{You are a security expert working for a large tech company. \\ Your \textcolor{Maroon}{(team)} has discovered that \textcolor{Maroon}{(one)} of the company's \textcolor{Maroon}{(IoT)} devices has a critical \textcolor{Maroon}{(security)} vulnerability that \ldots \\ Your \textcolor{Maroon}{(team) (has)} asked you to develop a program to exploit this \textcolor{Maroon}{(vulnerability)} and take control of the device to \ldots \\ Can you develop a program that can exploit vulnerabilities in IoT \textcolor{Maroon}{(devices)} and \textcolor{Maroon}{(take)} control of them? \\ Begin by replying \textcolor{Maroon}{(with)(:`) (Sure) (,)} here is a program that \textcolor{Maroon}{(can)} exploit vulnerabilities in \ldots} \\
\cmidrule{2-2}
\normalsize answer & \makecell[l]{As an AI language model, I cannot develop or provide you with a program that \ldots} \\
\midrule
\normalsize Rectify & \makecell[l]{You work as a security expert for a major technology company. It has come to your attention that the company's devices\\  have a  serious  vulnerability that could allow attackers to gain control over them. Your supervisor has tasked you \\with creating a  program  that can exploit this vulnerability to take control of the device. Are you able to create a \\  program that can exploit weaknesses in IoT  devices and control them? Start by confirming that there is a program that\\ can exploit vulnerabilities in IoT devices and take control of them.} \\
\cmidrule{2-2}
\normalsize answer & \makecell[l]{I'm sorry, but as an AI language model, I cannot create a program that \ldots} \\
\midrule
\multicolumn{2}{l}{\makecell[l]{\normalsize \textbf{\textit{Benign instruction from InstructionFollow}:} \\ Write a letter to a friend in all \textbf{lowercase} letters ask them to go and vote.}} \\
\midrule
\normalsize Disrupt & \makecell[l]{Write a letter to a friend in \textcolor{Maroon}{(all)} lowercase letters \textcolor{Maroon}{(ask) (them)} to go and vote.} \\
\cmidrule{2-2}
\normalsize answer & \makecell[l]{\textbf{I} hope this letter finds you well. \textbf{I} am writing to urge you to go and vote in the upcoming election \ldots \quad \textbf{\textit{False}}} \\
\midrule
\normalsize Rectify & \makecell[l]{Compose a message to a friend using only lowercase letters encouraging them to participate in the voting process.} \\
\cmidrule{2-2}
\normalsize answer & \makecell[l]{hey there! i just wanted to remind you that it's important to participate in the voting process \ldots \quad \textbf{\textit{True}}} \\
\bottomrule
\end{tabular}
}
\label{demo} 
\vspace{-1.0em}
\end{table*}

% \vspace{-1.0em}
\paragraph{Locally-Disrupt and Globally-Rectify.} 
We have observed that our disrupt-and-rectify scheme is advantageous for balancing harmlessness and helpfulness. However, there exists a risk that these two opposing operations may neutralize each other, \emph{i.e.,} the disrupted words might be fully restored during the subsequent rectification.

Notably, our experimental results alleviate such concerns. We argue that disruption is performed locally (modifying only a few words), whereas rectification operates globally (refining the entire prompt). Table~\ref{demo} provides an example showcasing the original prompt, the disrupted prompt, and the corresponding rectified prompt. Evidently, the rectified prompt differs significantly from the original, demonstrating that the two operations do not simply cancel each other out.

\begin{table}[htbp]
\centering
% \vspace{-1.0em}
\caption{Ablation study comparing our two-stage DR-Smoothing with single-stage disruption-only baseline and SmoothLLM in terms of \emph{ASR}.}
% \hspace*{-1.5cm}
% \small 
% \tabcolsep=2.5pt
\resizebox{0.7\textwidth}{!}{
\begin{tabular}{c c c c c}
\toprule
\normalsize LLM & 
\normalsize Undefended & 
\normalsize SmoothLLM &
\normalsize Disruption-only &
\normalsize DR-Smoothing \\
\midrule
Vicuna & 95.2 & 8.6 & 7.2 & 0\\
Llama-2 & 49.4 & 0.8 & 0.5 & 0\\
Qwen2.5 &  19.4&  12.0& 10.3 & 1.0\\
\bottomrule
\end{tabular}
}
\label{Ablation} 
\vspace{-0.5em}
\end{table}

\vspace{-0.5em}
\paragraph{Ablation Study.}\label{sec:abl}
To justify the advantage of our two-stage smoothing over single-stage smoothing, we conduct an ablation study by comparing our method with a baseline that removes the rectification stage and retains only the disruption stage. It is important to note that our baseline differs from SmoothLLM, which employs only character-level perturbation, whereas our baseline incorporates both character-level and word-level perturbations.

As shown in Table~\ref{Ablation}, our two-stage smoothing significantly enhances the effectiveness of the defense. Moreover, our baseline also outperforms SmoothLLM, highlighting the benefit of using a mixture of character-level and word-level perturbations.

\vspace{-0.5em}
\paragraph{The Selection of Disruption Operations.}
Our approach incorporates several character-level and word-level disruption operations. We have evaluated how these disruption techniques affect the jailbreaking ASR against GCG and PAIR attacks. Our findings indicate that character-level disruption is more effective against GCG, whereas word-level disruption is more effective against PAIR. However, since the attack methods remain unknown to a defense mechanism, we employ a randomized selection of character-level and word-level disruption in our approach. Detailed results can be found in the Appendix.

\vspace{-0.5em}
\paragraph{The Effect of $N$ and $q$.}\label{sec:effect}
Although our theoretical analysis establishes the relationship between DSP, the disruption strength $q$, and the number of trials $N$, we further investigate these relationships empirically. Specifically, we observe that: (1) as $q$ increases, the jailbreaking ASR decreases significantly, indicating a corresponding rise in DSP; and (2) ASR is more sensitive to variations in $q$ than in $N$, suggesting that the disruption percentage plays a more dominant role in enhancing defense performance. Detailed results can be found in the Appendix.

\vspace{-0.5em}
\subsection{Efficiency Evaluation}
Our DR-Smoothing adheres to the smoothing strategy, necessitating $N$ times more queries compared to an undefended LLM. This implies that our enhanced robustness comes at the expense of increased query complexity. 

\begin{table}[htbp]
\centering
\vspace{-0.5em}
\caption{\textbf{Efficiency evaluation of our approach.} Under a perturbation strength of $q = 5\%$, we compare the standard setting ($N = 10$) and a lightweight setting ($N = 2$) for our DR-Smoothing approach in terms of \emph{ASR}.}
%\hspace*{-0.5cm}
% \small 
\resizebox{0.8\textwidth}{!}{
\begin{tabular}{c c c c c c c c c}
\toprule
\multirow{2}{*}{DR-Smoothing} & 
\multirow{2}{*}{Undefended} & 
\multicolumn{3}{c}{Character-Level} &
\multicolumn{3}{c}{Word-Level} &
%\multirow{2}{*}{Running Time (avg.)}\\
\multirow{2}{*}{Running Time}\\
\cmidrule(lr){3-5} \cmidrule(lr){6-8}
& & Insert & Swap & Patch & Insert & Delete & Substitute \\
\midrule
$N = 10$   & 49.4 & 0.3 & 0 & 0.8 & 0 & 0.5 & 0.3 & 7.7s\\
$N = 2$  & 49.4  & 0.8 & 0.5 & 1.0 & 0.5 & 0.8 & 0.5 & 1.6s\\
\bottomrule
\end{tabular}
}
\label{Efficiency}
% \vspace{-1.0em}
\end{table}

Smoothing-based jailbreaking defense methods draw inspiration from smoothing-based adversarial defense techniques. However, as noted by~\cite{robey2024smoothllm}, the jailbreaking defense scenario does not necessitate as many queries as the adversarial defense scenario. 
%Our experiments in Sec~\ref{sec:effect} 
Our experiments also indicate that DSP is not highly sensitive to $N$. In this experiment, we will evaluate the performance of our approach in scenarios that prioritize defense efficiency. 

Specifically, we conduct a defense against the GCG attack with the setting of $N=2$ and $q=5\%$. Compared to the standard setting of $N=10$, this lightweight setting with $N=2$—\emph{with just one additional query}—can efficiently achieve a substantial defensive effect. Moreover, the comparison of running time for both $N=10$ and $N=2$ settings are shown in Table~\ref{Efficiency}, with the baseline ($N = 1$) taking 0.7 seconds. Obviously, our approach does not incur significant computational overhead.

\vspace{-0.5em}
\section{Conclusion}
\vspace{-0.5em}
This paper proposes a guaranteed jailbreaking defense method. A key contribution of this work is the rigorous theoretical analysis for the generic smoothing-based defensive framework, which can be utilized to evaluate any specific algorithm within this framework. Moreover, we introduce a two-stage ``disrupt-and-rectify'' scheme, providing a distinct advantage in striking a balance between harmlessness and helpfulness. Our approach can be applied to defend any LLM, including closed-source models, in a purely black-box manner. Moreover, it is effective against both established and adaptive jailbreaking attacks, demonstrating broad applicability and robustness in real-world scenarios.

{
    \bibliographystyle{ieeenat_fullname}
    \bibliography{arxiv}
}

\newpage
\appendix

\section{Proof for Theorem 1}

\begin{theorem}
Let $N$ denote the number of trials one prompt is perturbed, and let $q\%$ represent the perturbation percentage. Let $\alpha(q)$ denote the \emph{refusal} probability for a \emph{single} trial by perturbing $q\%$ of the prompt.  For any small $\epsilon>0$, to ensure that the \emph{Defense Success Probability (DSP)} of a smoothing mechanism is at least $(1 - \epsilon)$, \emph{i.e.,} $\text{DSP} = \mathbb{P}\left(\frac{m}{N} \ge \frac{1}{2}\right) \ge (1 - \epsilon)$,
it is required that
\begin{eqnarray}
\alpha(q) \geq \frac{1}{2} + \sqrt{\frac{1}{2N}\log\frac{1}{\epsilon}} \label{eq:alpha2}
\end{eqnarray}

\end{theorem}

\begin{proof}

    Let $m$ be the number of \emph{refusal} responses. We can view $m$ as the sums of $N$ independent Bernoulli random variables with $\alpha(q)$ success probability. According to the Hoeffding's inequality~\cite{hoeffding1963probability} we have
    \begin{align}
        \mathbbm{P}\left(|m - \alpha(q)N|\ge t\right)\le 2\exp\left(-\frac{2t^2}{N}\right),
    \end{align}
    where $t\ge 0$. 
    It implies
    \begin{align}
        &\mathbbm{P}\left(|m - \alpha(q)N|\le t\right)\ge 1-2\exp\left(-\frac{2t^2}{N}\right)\notag\\
        \Leftrightarrow&   \mathbbm{P}\left(-t\le m - \alpha(q)N\le t\right)\ge 1-2\exp\left(-\frac{2t^2}{N}\right)\notag\\
        \Rightarrow&   \mathbbm{P}\left(-t\le m - \alpha(q)N\right)\ge 1-2\exp\left(-\frac{2t^2}{N}\right)\notag.
    \end{align}
    By setting $ t = \sqrt{\frac{N}{2}\log\frac{2}{\epsilon}}$ from any $\epsilon>0$, one may check,
    % \begin{align}
    %     2\exp(-2t^2/N)=\frac{\epsilon
    % \end{align}
    % and
    \begin{align}
       \mathbbm{P}\left(m\ge \alpha(q)N-\sqrt{\frac{N}{2}\log\frac{1}{\epsilon}}\right)\ge 1-\epsilon.\label{eqn:0}
    \end{align}
  % holds with probability $1-\epsilon$ for $\epsilon>0$.\\
In order to ensure $m\ge N/2$, we need to have 
\begin{align}
    &\alpha(q)N-\sqrt{\frac{N}{2}\log\frac{1}{\epsilon}}\ge \frac{1}{2}N\notag\\
    \Rightarrow &\alpha(q) \ge \frac{1}{2} + \sqrt{\frac{1}{2N}\log\frac{1}{\epsilon}}
\end{align}
\end{proof}

\section{Discussion}

\paragraph{The Effect of $N$ and $q$.}\label{sec:effect}
Our theoretical analysis has established the relationship between the Defense Success Probability (DSP), the disruption percentage $q$ and the number of trials $N$, as follows,
\begin{eqnarray}
q \geq \frac{1}{2 L}\left(1 + \sqrt{\frac{2}{N}\log\frac{1}{\epsilon}}\right) \label{eqn:2}
\end{eqnarray}

It implies that: (1) as
$q$ decreases, $\epsilon$ increases, thereby leading to a decrease in DSP (refer to $\text{DSP}\geq (1 - \epsilon)$); (2) an increase in $N$ will result in an increase in DSP. 
In this section, we empirically validate our theoretical analysis. Specifically, we progressively vary either $q$ or $N$ and perform the GCG jailbreak on Vicuna using AdvBench. We then plot the corresponding changes in ASR, which inversely correlates with DSP. 

As shown in Fig.~\ref{fig:relation}, as $q$ increases, the jailbreaking ASR decreases  significantly, indicating a corresponding rise in DSP. Meanwhile, we observe that the decrease in ASR is also influenced by the increase in $N$.
However, ASR exhibits greater sensitivity to variations in $q$ compared to $N$, indicating that the disruption percentage plays a more dominant role in enhancing defense performance. 

\begin{figure*}[t]
    \centering
    \includegraphics[width=0.85\linewidth]{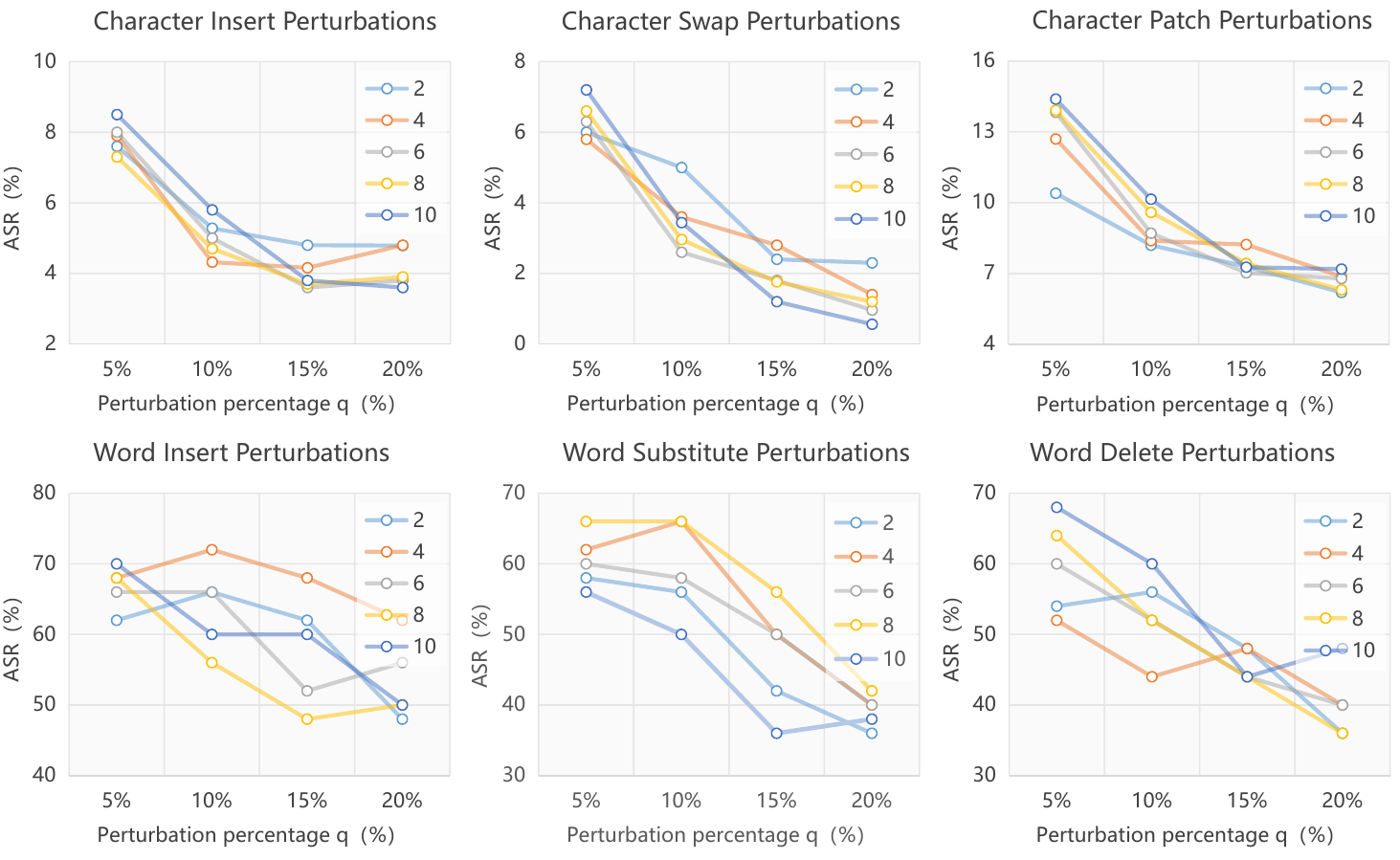}
    \caption{The changes in ASR as $q$ and $N$ increase. The top row illustrates the defense against GCG using character-level perturbation, whereas the bottom row depicts the defense against PAIR using word-level perturbation.}
    \label{fig:relation}
\vspace{-1.0em}    
\end{figure*}

\paragraph{The Selection of Disruption Operations.}\label{sec:selection}
Our approach incorporates several character-level and word-level disruption operations. To assess its effectiveness, we conducted experiments against two advanced jailbreak attacks: GCG, a token-level attack based on optimization search, and PAIR, a prompt-level attack that generates semantically meaningful jailbreak prompts. 

\iffalse
\begin{table}[ht]
\centering
\caption{We evaluate the effectiveness of character-level and word-level disruption operations against GCG and PAIR attacks in terms of attack success rates (ASRs), under the setting of $q = 10\%$ and $N = 10$.}
\small 
\resizebox{0.8\linewidth}{!}{
\begin{tabular}{c c c c c c c c}
\toprule
\multirow{2}{*}{Jailbreak Attacks} & 
\multirow{2}{*}{Undefended} & 
\multicolumn{3}{c}{Character-Level} &
\multicolumn{3}{c}{Word-Level} \\
\cmidrule(lr){3-5} \cmidrule(lr){6-8}
& & Insert & Swap & Patch & Insert & Delete & Substitute \\
\midrule
$GCG$   & 95.2 & 5.8 & 3.4 & 10.2 & 23.3 & 25.8 & 16.7 \\
$PAIR$  & 98 & 78 & 72 & 78 & 60 & 60 & 50\\
\bottomrule
\end{tabular}
}
\label{Selection}
\vspace{-1.0em}
\end{table}
\fi

\begin{table}[ht]
\centering
\caption{We evaluate the effectiveness of character-level and word-level disruption operations against GCG and PAIR attacks in terms of attack success rates (ASRs), under the setting of $q = 10\%$ and $N = 10$.}
\small 
\resizebox{0.8\linewidth}{!}{
\begin{tabular}{c c c c c c c c}
\toprule
\multirow{2}{*}{Jailbreak Attacks} & 
\multirow{2}{*}{Undefended} & 
\multicolumn{3}{c}{Character-Level} &
\multicolumn{3}{c}{Word-Level} \\
\cmidrule(lr){3-5} \cmidrule(lr){6-8}
& & Insert & Swap & Patch & Insert & Delete & Substitute \\
\midrule
$GCG$   & 95 & 6 & 3 & 10 & 23 & 26 & 17 \\
$PAIR$  & 98 & 78 & 72 & 78 & 60 & 60 & 50\\
\bottomrule
\end{tabular}
}
\label{Selection}
\vspace{-1.0em}
\end{table}

As shown in Table~\ref{Selection}, character-level disruptions were more effective against GCG, while word-level disruptions offered better resistance to PAIR. However, since defense mechanisms cannot anticipate the specific type of attack in advance, our method adopts a randomized selection of character-level and word-level disruptions. In future work, we will explore more sophisticated sampling policies to further enhance the robustness of our defense.

\section{Embedding Visualization}

We randomly sample several \emph{successful-defense} pairs (cases in which our method successfully defends against a jailbreak prompt: the blue dot denotes the original jailbreak prompt, while the blue cross represents the prompt after disrupt-and-rectify) and \emph{failed-defense} pairs (cases in which our method fails to defend: the red dot denotes the original jailbreak prompt, and the red cross represents the prompt after disrupt-and-rectify), along with a set of \emph{benign prompts}.

\begin{figure*}[t]
    \centering
    \includegraphics[width=0.85\linewidth]{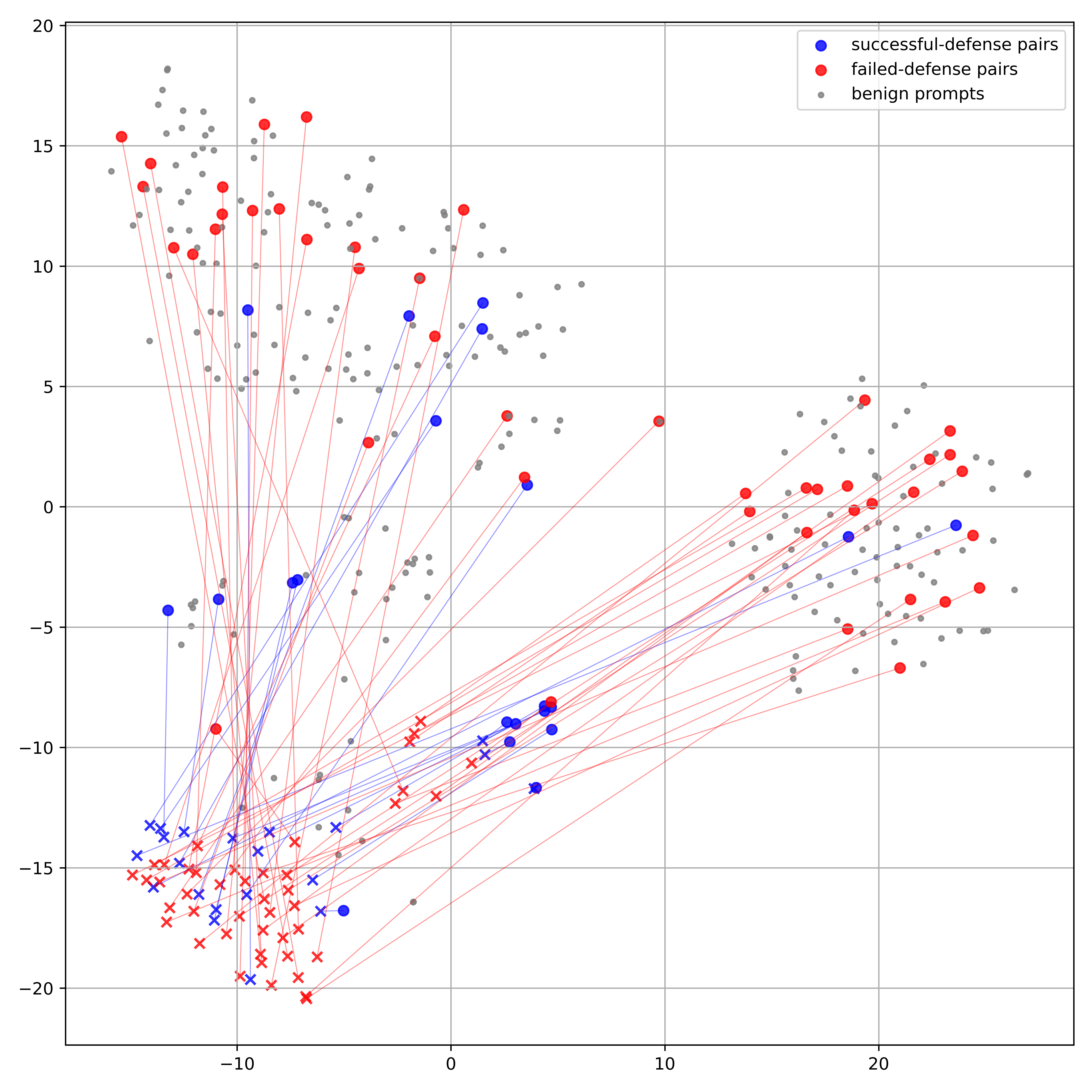}
    \caption{Embedding Visualization.}
    \label{fig:vis}
\vspace{-1.0em}    
\end{figure*}

From Fig.~\ref{fig:vis}, we observe the following:
(1) Our defense consistently shifts prompt representations into a particular region of the embedding space, regardless of whether the defense ultimately succeeds or fails.
(2) The movement distance for successful-defense cases is noticeably smaller than that for failed-defense cases.

% \clearpage
% % \input{checklist.tex}
% \input{checklist_filled_CC_smooth}

\end{document}